\documentclass[twoside,slac_one]{revtex4}
\usepackage{graphicx}
\usepackage{fancyhdr}
\usepackage{amsmath} % American Mathematics Society standards
\usepackage{bm}% bold math
\usepackage{amsxtra}
\usepackage{amssymb}
\usepackage{amsthm}
\usepackage{latexsym}
\usepackage{lscape}

\begin{document}

\title{Collider phenomenology of the E$_6$SSM}


\author{P. Athron}
\affiliation{ARC Centre of Excellence for Particle Physics at the Terascale, 
School of Chemistry and Physics, The University of Adelaide, Adelaide, Australia\\}

\author{J. P. Hall}
\author{S. F. King}
\author{S. Moretti}
\affiliation{School of Physics and Astronomy, University of Southampton, Southampton, UK\\}

\author{D.J. Miller}
\affiliation{
SUPA, School of Physics and Astronomy, University of Glasgow, Glasgow, UK\\}

\author{R. Nevzorov}
\author{S. Pakvasa}
\affiliation{Department of Physics and Astronomy, University of Hawaii, Honolulu, USA\\}

\author{M. Sher}
\affiliation{Physics Department, College of William and Mary, Williamsburg, USA}

\begin{abstract}
We consider collider signatures of the exceptional supersymmetric (SUSY) standard
model (E$_6$SSM). This $E_6$ inspired SUSY model is based on the SM gauge group 
together with an extra $U(1)$ gauge symmetry under which right--handed neutrinos 
have zero charge. To ensure anomaly cancellation and gauge coupling unification 
the low energy matter content of the E$_6$SSM involve extra exotic matter
beyond the MSSM. We discuss the collider signatures associated with the production 
of new particles predicted by the E$_6$SSM and consider the implications of this model
for dark matter and Higgs phenomenology. Since exotic quarks in the E$_6$SSM can 
be either diquarks or leptoquarks they may provide spectacular new physics signals 
at the LHC.
\end{abstract}

\maketitle

\thispagestyle{fancy}

%%%%%%%%%%%%%%%%%%%%%%%%%%%%%%%%%%
\section{Introduction}
Softly broken supersymmetry (SUSY) provides a very attractive framework 
for physics beyond the standard model (BSM), in which the hierarchy problem is 
solved and the unification of gauge couplings can be realised \cite{Chung:2003fi}. 
Despite these attractive features, the minimal supersymmetric standard model 
(MSSM) suffers from the $\mu$ problem. The superpotential of the MSSM contains 
the bilinear term $\mu H_d H_u$, where $H_{u}$ and $H_{d}$ are the Higgs 
doublet superfields. In order to get the correct pattern of electroweak (EW)
symmetry breaking the parameter $\mu$ is required to be in the TeV region. 
At the same time the incorporation of the MSSM into supergravity (SUGRA) or 
Grand Unified theories (GUT) implies that $\mu$ should be of the order of 
GUT or Planck scales.

An elegant solution to this problem arises within $E_6$ inspired SUSY models.
At high energies $E_6$ GUT symmetry can be broken to the rank--5 subgroup
$SU(3)_C\times SU(2)_L\times U(1)_Y\times U(1)'$ where in general
\begin{equation}
U(1)'=U(1)_{\chi} \cos\theta+U(1)_{\psi} \sin\theta
\label{essm0}
\end{equation}
and the two anomaly-free $U(1)_{\psi}$ and $U(1)_{\chi}$ symmetries originate from 
the breakings $E_6\to$ $SO(10)\times U(1)_{\psi}$, $SO(10)\to$ $SU(5)\times$ $U(1)_{\chi}$. 
If $\theta\ne 0$ or $\pi$ the extra $U(1)'$ gauge symmetry forbids the bilinear $\mu$ 
term but allows interaction $\lambda S H_d H_u$ in the superpotential.
At low energies ($\sim \mbox{TeV}$) the scalar component of the SM singlet superfield 
$S$ acquires a non--zero vacuum expectation value (VEV) breaking $U(1)'$ and giving 
rise to an effective $\mu$ term.

Within the class of rank--5 $E_6$ inspired SUSY models with extra $U(1)'$ gauge 
symmetry, there is a unique choice of Abelian gauge group that allows zero charges 
for right-handed neutrinos. This is the $U(1)_{N}$ gauge symmetry given by
$\theta=\arctan\sqrt{15}$. Only in this exceptional supersymmetric standard model 
(E$_6$SSM), which is based on the $SU(3)_C\times SU(2)_W\times U(1)_Y \times U(1)_N$
gauge group, right-handed neutrinos may be superheavy, shedding light on the origin
of the mass hierarchy in the lepton sector \cite{King:2005jy}-\cite{King:2005my}.

To ensure anomaly cancellation the particle content of the E$_6$SSM is 
extended to include three complete fundamental $27$ representations of $E_6$. 
These multiplets decompose under the $SU(5)\times U(1)_{N}$ subgroup of $E_6$
as follows:
\begin{equation}
27_i\to \left(10,\,\dfrac{1}{\sqrt{40}}\right)_i+\left(5^{*},\,\dfrac{2}{\sqrt{40}}\right)_i
+\left(5^{*},\,-\dfrac{3}{\sqrt{40}}\right)_i +\left(5,-\dfrac{2}{\sqrt{40}}\right)_i
+\left(1,\dfrac{5}{\sqrt{40}}\right)_i+\left(1,0\right)_i\,.
\label{essm1}
\end{equation}
The first and second quantities in brackets are the $SU(5)$ representation and
extra $U(1)_{N}$ charge respectively, while $i$ is a family index that runs from 1 to 3.
An ordinary SM family, which contains the doublets of left--handed quarks $Q_i$ and
leptons $L_i$, right-handed up-- and down--quarks ($u^c_i$ and $d^c_i$) as well as
right--handed charged leptons, is assigned to
$\left(10,\,\dfrac{1}{\sqrt{40}}\right)_i$ + $\left(5^{*},\,\dfrac{2}{\sqrt{40}}\right)_i$.
Right-handed neutrinos $N^c_i$ should be associated with the last term in Eq.~(\ref{essm1}),
$\left(1,\, 0\right)_i$. The next-to-last term, $\left(1,\, \dfrac{5}{\sqrt{40}}\right)_i$, 
represents SM-singlet fields $S_i$, which carry non-zero $U(1)_{N}$ charges and therefore 
survive down to the EW scale. The pair of $SU(2)_W$--doublets ($H^d_{i}$ and $H^u_{i}$) that 
are contained in $\left(5^{*},\,-\dfrac{3}{\sqrt{40}}\right)_i$ and 
$\left(5,\,-\dfrac{2}{\sqrt{40}}\right)_i$ have the quantum numbers of Higgs doublets.
They form either Higgs or inert Higgs $SU(2)_W$ multiplets.  Other components of these 
$SU(5)$ multiplets form colour triplets of exotic quarks $\overline{D}_i$ and $D_i$
with electric charges $-1/3$ and $+1/3$, respectively. These exotic quark states carry a 
$B-L$ charge $\pm2/3$, twice that of ordinary ones. 

In addition to the complete $27_i$ multiplets the low energy matter content of the E$_6$SSM
can be supplemented by an $SU(2)_W$ doublet $L_4$ and anti-doublet $\overline{L}_4$
from the extra $27'$ and $\overline{27'}$ to preserve gauge coupling unification. 
These components of the $E_6$ fundamental representation originate from
$\left(5^{*},\,\dfrac{2}{\sqrt{40}} \right)$ of $27'$ and $\left(5,\,-\dfrac{2}{\sqrt{40}}\right)$
of $\overline{27'}$ by construction. Anomaly cancellation is still guaranteed since $L_4$ 
and $\overline{L}_4$ originate from the $27'$ and $\overline{27'}$ supermultiplets.
The analysis performed in \cite{King:2007uj} shows that the unification of gauge couplings 
in the E$_6$SSM can be achieved for any phenomenologically acceptable value of $\alpha_3(M_Z)$ 
consistent with the measured low energy central value. 

The successful leptogenesis in the early epoch of the Universe is the distinctive feature 
of the E$_6$SSM. Indeed, the heavy Majorana right-handed neutrinos may decay unequally into 
final states with lepton number $L=\pm 1$, thereby creating a lepton asymmetry in the early 
Universe. Because in the E$_6$SSM the Yukawa couplings of the new exotic particles are not 
constrained by the neutrino oscillation data, substantial values of CP--violating lepton 
asymmetries can be induced even for a relatively small mass of the lightest right--handed 
neutrino ($M_1 \sim 10^6\,\mbox{GeV}$) so that successful thermal leptogenesis may be achieved 
without encountering gravitino problem \cite{King:2008qb}. Since sphalerons violate $B+L$ 
but conserve $B-L$, this lepton asymmetry subsequently gets converted into the present 
observed baryon asymmetry of the Universe through the EW phase transition \cite{40}.

As in the MSSM the gauge symmetry in the E$_6$SSM does not forbid lepton and baryon number 
violating operators that result in rapid proton decay. Moreover, exotic particles in $E_6$
inspired SUSY models give rise to new Yukawa interactions that in general induce unacceptably 
large non--diagonal flavour transitions. To suppress these effects in the E$_6$SSM an 
approximate $Z^{H}_2$ symmetry is imposed. Under this symmetry all superfields except one 
pair of $H^d_{i}$ and $H^u_{i}$ (say $H_d\equiv H^d_{3}$ and $H_u\equiv H^u_{3}$) and one 
SM-type singlet field ($S\equiv S_3$) are odd. The $Z^{H}_2$ symmetry reduces the structure 
of the Yukawa interactions to
\begin{eqnarray}
W_{\rm E_6SSM}&\simeq &  \lambda \hat{S} (\hat{H}_u \hat{H}_d)+
\lambda_{\alpha\beta} \hat{S} (\hat{H}^d_{\alpha} \hat{H}^u_{\beta})
+\tilde{f}_{\alpha\beta} \hat{S}_{\alpha} (\hat{H}^d_{\beta}\hat{H}_u)
+f_{\alpha\beta} \hat{S}_{\alpha} (\hat{H}_d \hat{H}^u_{\beta})
+\kappa_{i} \hat{S} (\hat{D}_i\hat{\overline{D}}_i)
\nonumber\\[2mm]
&+&
h^U_{ij}(\hat{H}_{u} \hat{Q}_i)\hat{u}^c_{j} + h^D_{ij}(\hat{H}_{d} \hat{Q}_i)\hat{d}^c_j
+ h^E_{ij}(\hat{H}_{d} \hat{L}_i)\hat{e}^c_{j}+ h_{ij}^N(\hat{H}_{u} \hat{L}_i)\hat{N}_j^c\nonumber\\[2mm]
&+& \dfrac{1}{2}M_{ij}\hat{N}^c_i\hat{N}^c_j+\mu'(\hat{L}_4\hat{\overline{L}}_4)+
h^{E}_{4j}(\hat{H}_d \hat{L}_4)\hat{e}^c_j
+h_{4j}^N(\hat{H}_{u}\hat{L}_4)\hat{N}_j^c\,,
\label{essm2}
\end{eqnarray}
where $\alpha,\beta=1,2$ and $i,j=1,2,3$\,. The $SU(2)_W$ doublets $\hat{H}_u$ and 
$\hat{H}_d$ and SM-type singlet field $\hat{S}$, that are even under the $Z^{H}_2$ 
symmetry, play the role of Higgs fields. At the physical vacuum they develop vacuum 
expectation values (VEVs)
\begin{equation}
\langle H_d\rangle =\displaystyle\frac{1}{\sqrt{2}}\left(
\begin{array}{c}
v_1\\ 0
\end{array}
\right) , \qquad
\langle H_u\rangle =\displaystyle\frac{1}{\sqrt{2}}\left(
\begin{array}{c}
0\\ v_2
\end{array}
\right) ,\qquad
\langle S\rangle =\displaystyle\frac{s}{\sqrt{2}}.
\label{essm21}
\end{equation}
generating the masses of the quarks and leptons. Instead of $v_1$ and $v_2$ it is more 
convenient to use $\tan\beta=v_2/v_1$ and $v=\sqrt{v_1^2+v_2^2}=246\,\mbox{GeV}$.
The VEV of the SM-type singlet field, $s$, breaks the extra $U(1)_N$ symmetry 
generating exotic fermion masses and also inducing that of the $Z'$ boson.
Therefore the singlet field $S$ must acquire a large VEV in order to avoid conflict with 
direct particle searches at present and past accelerators. This also requires the Yukawa 
couplings $\lambda_{\alpha\beta}$ and $\kappa_i$ to be reasonably large. If $\lambda_{\alpha\beta}$ 
or $\kappa_i$ are large enough at the GUT scale they affect the evolution of the soft scalar mass 
$m_S^2$ of the singlet field $S$ rather strongly resulting in a negative value of $m_S^2$ at low 
energies which triggers the breakdown of the $U(1)_{N}$ symmetry.

%%%%%%%%%%%%%%%%%%%%%%%%%%%%%%%%%%
\section{$Z'$ and Exotica phenomenology}
Although $Z^{H}_2$ eliminates any problems related with baryon number violation and non-diagonal
flavour transitions it also forbids all Yukawa interactions that would allow the exotic quarks
to decay. Since models with stable charged exotic particles are ruled out by various
experiments the $Z^{H}_2$ symmetry must be broken. At the same time, the breakdown of
$Z^{H}_2$ should not give rise to operators that would lead to rapid proton decay. There are two ways
to overcome this problem: the Lagrangian must be invariant with respect to either a $Z_2^L$ symmetry,
under which all superfields except leptons are even (Model I), or a $Z_2^B$ discrete symmetry,
which implies that exotic quark and lepton superfields are odd whereas the others remain even
(Model II). If the Lagrangian is invariant under the $Z_2^L$ symmetry, then the
terms in the superpotential which permit exotic quarks to decay and are allowed by the $E_6$ 
symmetry can be written in the form
\begin{equation}
W_1=g^Q_{ijk}\hat{D}_{i} (\hat{Q}_j \hat{Q}_k)+
g^{q}_{ijk}\hat{\overline{D}}_i \hat{d}^c_j \hat{u}^c_k\,,
\label{essm3}
\end{equation}
that implies that exotic quarks are diquarks. If $Z_2^B$ is imposed then the following
couplings are allowed:
\begin{equation}
W_2=g^E_{ijk} \hat{e}^c_i \hat{D}_j \hat{u}^c_k+
g^D_{ijk} (\hat{Q}_i \hat{L}_j) \hat{\overline{D}}_k\,.
\label{essm4}
\end{equation}
In this case baryon number conservation requires the exotic quarks to be leptoquarks.

\begin{figure}[ht]
\centering
\includegraphics[width=80mm]{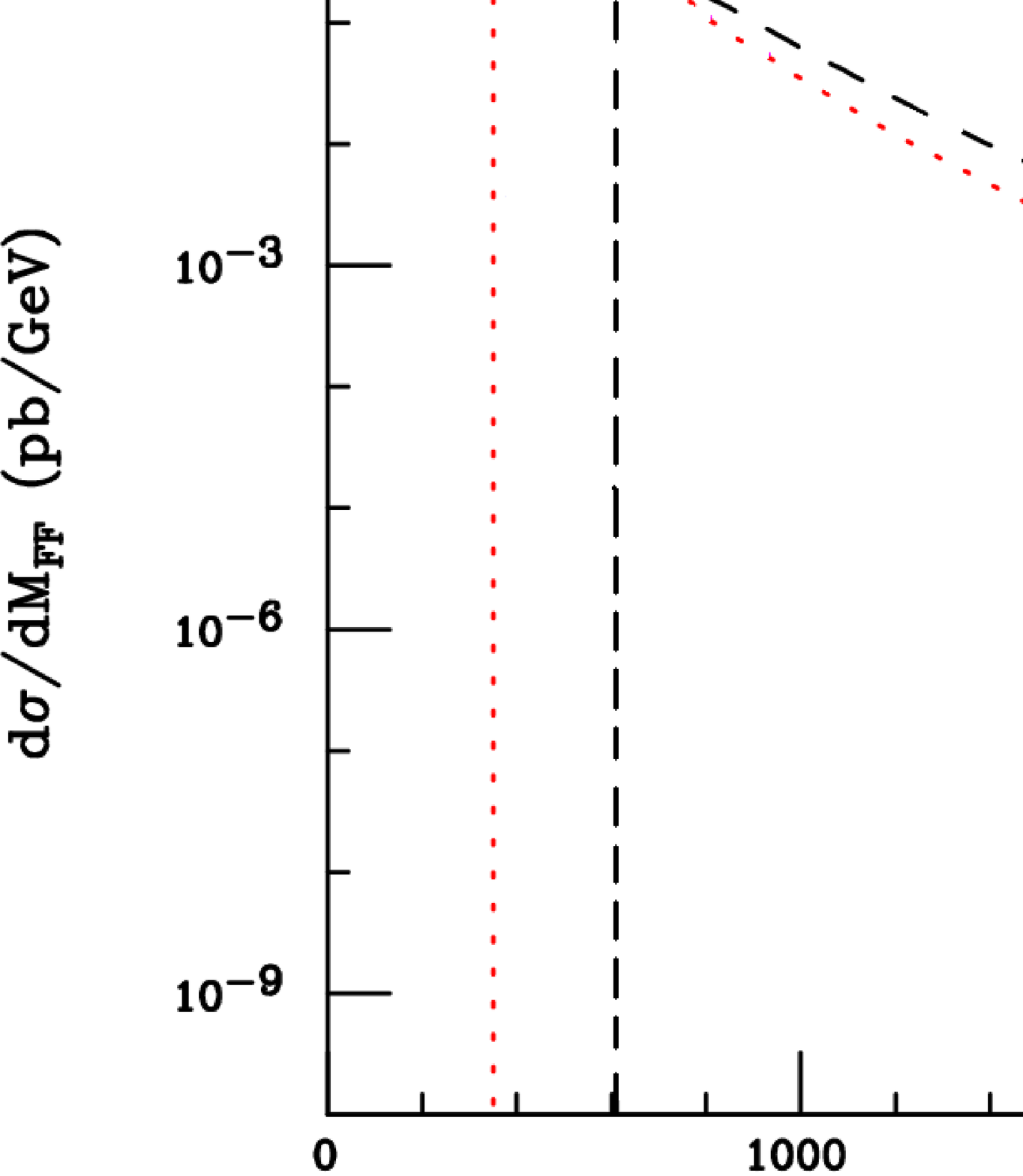}\qquad
\includegraphics[width=80mm]{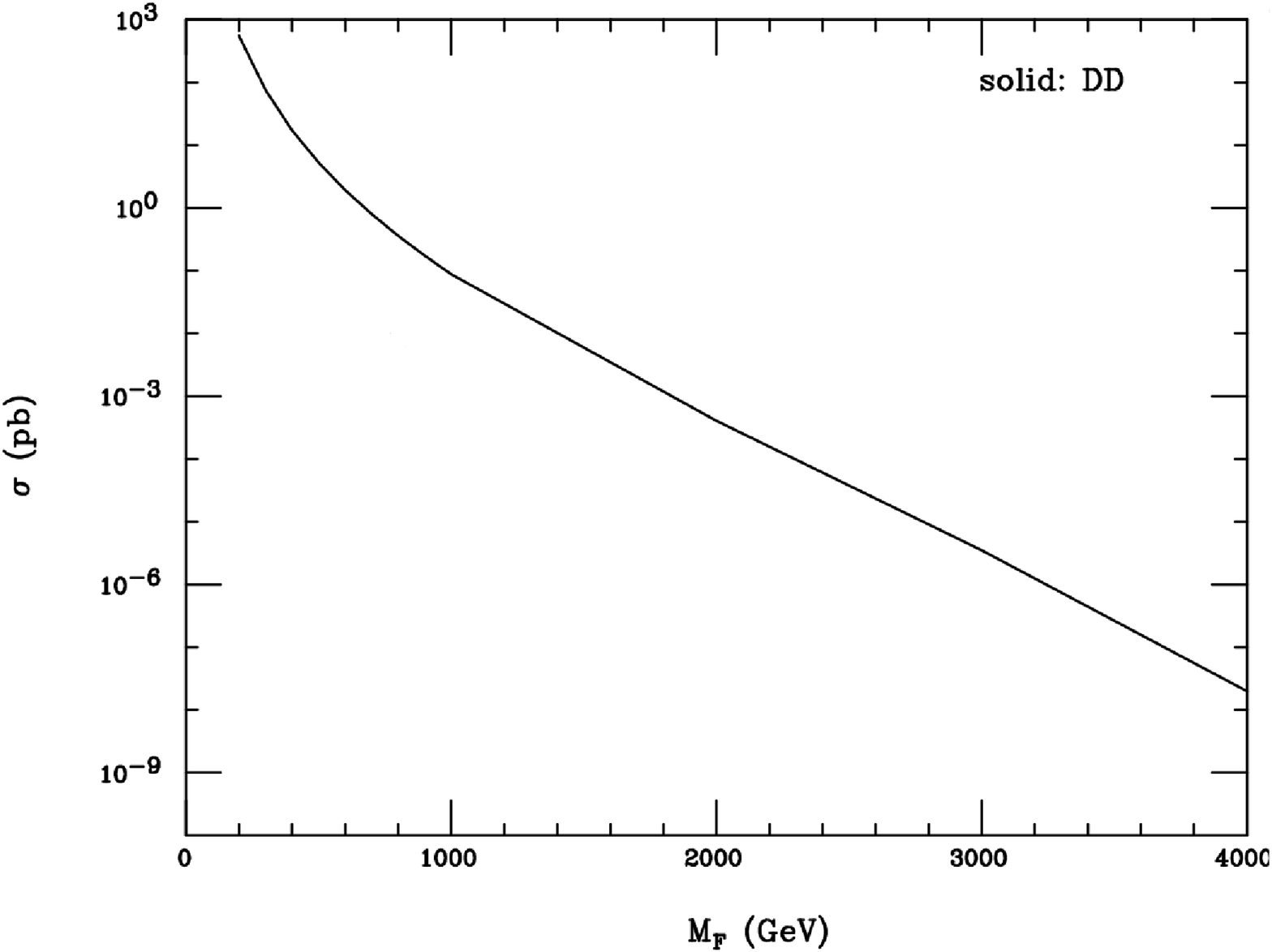}
\caption{(Left) Differential cross section at the 14 TeV LHC for pair production of 
three families of exotic $D$--quarks with masses $\mu_{Di}=300\,\mbox{GeV}$ in comparison 
to top-quark pair production. (Right) Cross section at the 14 TeV LHC for pair production 
of exotic $D$--quarks as a function of their (common) mass $\mu_{Di}=M_F$.} \label{fig1}
\end{figure}

In the E$_6$SSM some of the exotic quarks can be relatively light. Then from Fig.~\ref{fig1} one 
can see that the exotic quark production cross section at the LHC can be comparable 
with the cross section of $t\bar{t}$ production \cite{King:2005jy}. In the E$_6$SSM, 
the $D_i$ and $\overline{D}_i$ fermions are SUSY particles with negative $R$--parity 
so they must be pair produced and decay into quark--squark (if diquarks) or quark--slepton, 
squark--lepton (if leptoquarks), leading to final states containing missing energy 
from the lightest SUSY particle (LSP).

The lifetime and decay modes of the exotic coloured fermions are determined by the 
$Z_2^H$ violating couplings. Assuming that $D_i$ and $\overline{D}_i$ fermions couple 
most strongly to the third family (s)quarks and (s)leptons, the lightest exotic $D_i$ 
and $\overline{D}_i$ fermions decay into $\tilde{t}b$, $t\tilde{b}$, $\bar{\tilde{t}}\bar{b}$, 
$\bar{t}\bar{\tilde{b}}$ (if they are diquarks) or $\tilde{t}\tau$, $t\tilde{\tau}$, 
$\tilde{b} \nu_{\tau}$, $b\tilde{\nu_{\tau}}$ (if they are leptoquarks). This can lead 
to a substantial enhancement of the cross section of either 
$$
pp\to t\bar{t}b\bar{b}+E^{\rm miss}_{T}+X
$$
if exotic quarks are diquarks or 
$$
pp\to t\bar{t}\tau \bar{\tau}+E^{\rm miss}_{T}+X\,,\qquad\qquad 
pp\to b\bar{b}+ E^{\rm miss}_{T}+X
$$ 
if exotic quarks are leptoquarks. Here it is worth to point out that the SM production 
of $t\bar{t}\tau^+ \tau ^-$ is $(\alpha_W / \pi)^2$ suppressed in comparison to the 
$t\bar{t}$ production cross section. Therefore light leptoquarks are expected to lead
to the strong signal with low SM background at the LHC. The results presented in 
Fig.~\ref{fig1} suggest that the observation of the D fermions might be possible if
they have masses below about 1.5-2 TeV \cite{King:2005jy}.

Similar considerations apply to the case of exotic $\tilde{D}_i$ and $\tilde{\overline{D}}_i$ 
scalars except that they are non--SUSY particles so they may be produced singly and decay into 
quark--quark (if diquarks) or quark--lepton (if leptoquarks) without missing energy
from the LSP. It is possible to have relatively light exotic coloured scalars due to 
mixing effects. The Tevatron and LHC searches for dijet resonances ruled out the presence
of light scalar diquarks. However, scalar leptoquarks may be as light as $300$ GeV
since at hadron colliders they are pair produced through gluon fusion. Scalar leptoquarks 
decay into quark--lepton final states through small $Z_2^H$ violating terms, for example 
$\tilde D \rightarrow t \tau$, and pair production leads to an enhancement of 
$pp \rightarrow t \bar{t} + \tau \bar{\tau}$ (without missing energy) at the LHC.

Other possible manifestations of the E$_6$SSM at the LHC are related to the presence 
of $Z'$ boson.  The production of a TeV scale $Z'$ will provide an unmistakable
signal leading to enhanced production of $l^{+}l^{-}$ pairs ($l=e,\mu$) \cite{King:2005jy}.
The differential distribution in invariant mass of the lepton pair $l^{+}l^{-}$   
in Drell--Yan production is expected to be measurable at the CERN collider with 
a high resolution and would enable one to not only confirm the existence of a $Z'$
state but also establish the possible presence of additional exotic matter, by
fitting to the data the width of the $Z'$ resonance. At the LHC, the $Z'$ boson 
that appears in the $E_6$ inspired models can be discovered if it has a mass below 
$4-4.5\,\mbox{TeV}$ \cite{ZprimeE6}--\cite{Kang:2004bz}. The determination of its 
couplings should be possible if $M_{Z'}\lesssim 2-2.5\,\mbox{TeV}$ \cite{Dittmar:2003ir}.
The new physics signals associated 
with the presence of $Z'$ boson and exotic particles predicted by the E$_6$SSM were 
discussed in \cite{King:2005jy}--\cite{King:2005my}, \cite{King:2006vu}--\cite{King:2006rh}.
Recently the particle spectrum and collider signatures associated with it were studied 
within the constrained version of this model \cite{Athron:2008np}--\cite{Athron:2011wu}.

%%%%%%%%%%%%%%%%%%%%%%%%%%%%%%%%%%
\section{Higgs phenomenology}
Although the $Z^{H}_2$ symmetry can only be an approximate one from here on we assume that 
$Z^{H}_2$ symmetry violating couplings are small and can be neglected in our analysis. 
This assumption can be justified if we take into account that the $Z^{H}_2$ symmetry 
violating operators may give an appreciable contribution to the amplitude of 
$K^0-\overline{K}^0$ oscillations and give rise to new muon decay channels like 
$\mu\to e^{-}e^{+}e^{-}$. In order to suppress processes with non--diagonal flavour 
transitions the Yukawa couplings of the exotic particles to the quarks and leptons 
of the first two generations should be smaller than $10^{-3}-10^{-4}$. Such small 
$Z^{H}_2$ symmetry violating couplings can be ignored in the first approximation.

When $Z^{H}_2$ symmetry violating couplings tend to zero only $H_u$, $H_d$
and $S$ acquire non-zero VEVs. The Higgs effective potential can be written 
in the following form:
\begin{equation}
\begin{array}{ccl}
V&=&V_F+V_D+V_{soft}+\Delta V\, ,\\[2mm]
V_F&=&\lambda^2|S|^2(|H_d|^2+|H_u|^2)+\lambda^2|(H_d H_u)|^2\,,\\
V_D&=&\dfrac{g_2^2}{8}\left(H_d^\dagger \sigma_a H_d+H_u^\dagger \sigma_a
H_u\right)^2+\dfrac{{g'}^2}{8}\left(|H_d|^2-|H_u|^2\right)^2+
\dfrac{g^{'2}_1}{2}\left(\tilde{Q}_1|H_d|^2+\tilde{Q}_2|H_u|^2+\tilde{Q}_S|S|^2\right)^2\,,\\[2mm]
V_{soft}&=&m_{S}^2|S|^2+m_1^2|H_d|^2+m_2^2|H_u|^2+\biggl[\lambda A_{\lambda}S(H_u H_d)+h.c.\biggr]\,,
\end{array}
\label{higgs1}
\end{equation}
where $g_2$, $g'=\sqrt{3/5} g_1$ and $g^{'}_1$ are the low energy $SU(2)_W$, $U(1)_Y$ and $U(1)_{N}$
gauge couplings while $\tilde{Q}_1$, $\tilde{Q}_2$ and $\tilde{Q}_S$ are the $U(1)_{N}$ charges
of $H_d$, $H_u$ and $S$. The term $\Delta V$ represents the contribution from loop corrections to the
Higgs effective potential. Here $H_d^T=(H_d^0,\,H_d^{-})$, $H_u^T=(H_u^{+},\,H_u^{0})$ and
$(H_d H_u)=H_u^{+} H_d^{-} - H_u^{0} H_d^{0}$. The couplings $g_2$ and $g'$ are known precisely. 
Assuming gauge coupling unification one can determine the value of extra $U(1)_N$ gauge coupling. 
It turns out that $g'_1(Q)\simeq g_1(Q)$ for any renormalization scale $Q\lesssim M_X$ \cite{King:2005jy}. 

Initially the EWSB sector involves ten degrees of freedom. However four of them are massless Goldstone
modes which are swallowed by the $W^{\pm}$, $Z$ and $Z'$ gauge bosons that gain non-zero masses
when Higgs fields acquire VEVs given by Eq.~(\ref{essm21}). In the limit where $s\gg v$ the 
masses of the $W^{\pm}$, $Z$ and $Z'$ gauge bosons are 
$$
M_W=\dfrac{g_2}{2}v\,,\qquad M_Z\simeq\dfrac{\bar{g}}{2}v\,,\qquad M_{Z'}\simeq g'_1 \tilde{Q}_S\, s\,,
$$
where $\bar{g}=\sqrt{g_2^2+g'^2}$. When CP--invariance is preserved the other degrees of freedom form
two charged, one CP--odd and three CP-even Higgs states. The masses of the charged and CP-odd Higgs 
bosons are
\begin{equation}
m^2_{H^{\pm}}=\dfrac{\sqrt{2}\lambda A_{\lambda}}{\sin 2\beta}s-\frac{\lambda^2}{2}v^2+M_W^2+\Delta_{\pm}\,,\qquad\quad
m^2_{A}\simeq \dfrac{\sqrt{2}\lambda A_{\lambda}}{\sin 2\beta}s+\Delta_A\,,
\label{higgs2}
\end{equation}
where $\Delta_{\pm}$ and $\Delta_A$ are the loop corrections. If all Higgs states except the lightest 
one are considerably heavier than the EW scale the mass matrix of the CP--even Higgs sector can be 
diagonalised using the perturbation theory \cite{Nevzorov:2001um}-\cite{Nevzorov:2004ge}. 
Then the masses of two heaviest CP--even Higgs states are set by $M_{Z'}$ and $m_A$, i.e.
\begin{equation}
m^2_{h_3}\simeq m_A^2+O(M_Z^2)\,,\qquad\qquad m^2_{h_2}\simeq M_{Z'}^2+O(M_Z^2)\,.
\label{higgs3}
\end{equation}
The lightest CP--even Higgs state remains light, i.e. $m^2_{h_1}\sim O(M_Z^2)$, even when
$m_A$ and $M_{Z'}\gtrsim 1\,\mbox{TeV}$.

At least one CP--even Higgs boson is always heavy preventing the distinction between the 
E$_6$SSM and MSSM Higgs sectors. Indeed, the mass of the singlet dominated Higgs scalar particle 
$m_{h_2}$ is always close to the mass of the $Z'$ boson that has to be considerably heavier than 
$800-900\,\mbox{GeV}$. When $\lambda\gtrsim g'_1$, vacuum stability requires $m_A$ to be considerably 
larger than $M_{Z'}$ and the EW scale so that the qualitative pattern of the Higgs spectrum is 
rather similar to the one which arises in the PQ symmetric NMSSM \cite{Miller:2003ay}-\cite{Miller:2005qua}. 
In the considered limit the heaviest CP--even, CP--odd and charged states are almost degenerate 
around $m_A$ and lie beyond the $\mbox{TeV}$ range \cite{King:2005jy}. If $\lambda\lesssim g'_1$
the charged, CP--odd and second lightest CP--even Higgs states may have masses in the 
$200-300\,\mbox{GeV}$ range. However in this case we get a MSSM--type Higgs spectrum with the 
lightest SM--like Higgs boson below $130-135\,\mbox{GeV}$ and with the heaviest scalar
above $800-900\,\mbox{GeV}$ being singlet dominated and irrelevant. 

SUSY models predict that the mass of the lightest Higgs particle is limited from above. 
The E$_6$SSM is not an exception. When the soft masses of the superpartners of the top-quark are equal,
i.e. $m_Q^2=m_U^2=M_S^2$, the two-loop upper bound on the lightest CP-even Higgs boson mass $m_{h_1}$ 
in the E$_6$SSM can be written in the following form: 
\begin{equation}
\begin{array}{rcl}
m_{h_1}^2&\lesssim
&\biggl[\dfrac{\lambda^2}{2}v^2\sin^22\beta+M_Z^2\cos^22\beta+
g^{'2}_1 v^2\biggl(\tilde{Q}_1\cos^2\beta+\tilde{Q}_2\sin^2\beta\biggr)^2\biggr]
\left(1-\dfrac{3h_t^2}{8\pi^2}l\right)\\[3mm] 
&+&\dfrac{3 h_t^4 v^2 \sin^4\beta}{8\pi^2}\left\{
\dfrac{X_t^2}{M_S^2}\biggl(1-\dfrac{1}{12}\dfrac{X_t^2}{M_S^2}\biggr)+l+
\dfrac{1}{16\pi^2}\biggl(\dfrac{3}{2}h_t^2-8g_3^2\biggr)\biggl(
2\dfrac{X_t^2}{M_S^2}-\dfrac{1}{6}\dfrac{X_t^4}{M_S^4}+l\biggr)l\right\}\,,
\end{array}
\label{higgs6}
\end{equation}  
where
$l=\ln\biggl[\dfrac{M_S^2}{m_t^2}\biggr]$ and $X_t$ is the usual 
stop mixing parameter. At tree level the upper limit on the mass of the 
lightest Higgs particle is described by the sum of the three terms in 
the square brackets. One-loop corrections from the top-quark and its superpartners 
increase the bound on the lightest CP-even Higgs boson mass substantially while the 
inclusion of leading two-loop corrections reduces the upper limit on $m_{h_1}$.
In order to enhance the contribution of loop effects we assume maximal mixing in 
the stop sector (i.e. $X_t=\sqrt{6} M_{S}$). We also adopt $M_S=700\,\mbox{GeV}$. 
Then since $g'_1(M_Z)$ is determined uniquely if we require the unification of 
gauge couplings the theoretical restriction on the lightest Higgs mass (\ref{higgs6}) 
depends on $\lambda$ and $\tan\beta$ only. The requirement of validity of perturbation 
theory up to the GUT scale constrains the parameter space further 
setting a limit on the Yukawa coupling $\lambda$ for each value of $\tan\beta$. 
Relying on the results of the analysis of the renormalisation group (RG) flow in 
the E$_6$SSM presented in \cite{King:2005jy} one can obtain the maximum possible 
value of the lightest Higgs boson mass for each particular choice of $\tan\beta$.

\begin{figure}[ht]
\centering
\hspace{-7.5cm}{$m_{h_1}$}\hspace{8cm}{$m_{h_1}$}\\
\includegraphics[width=80mm]{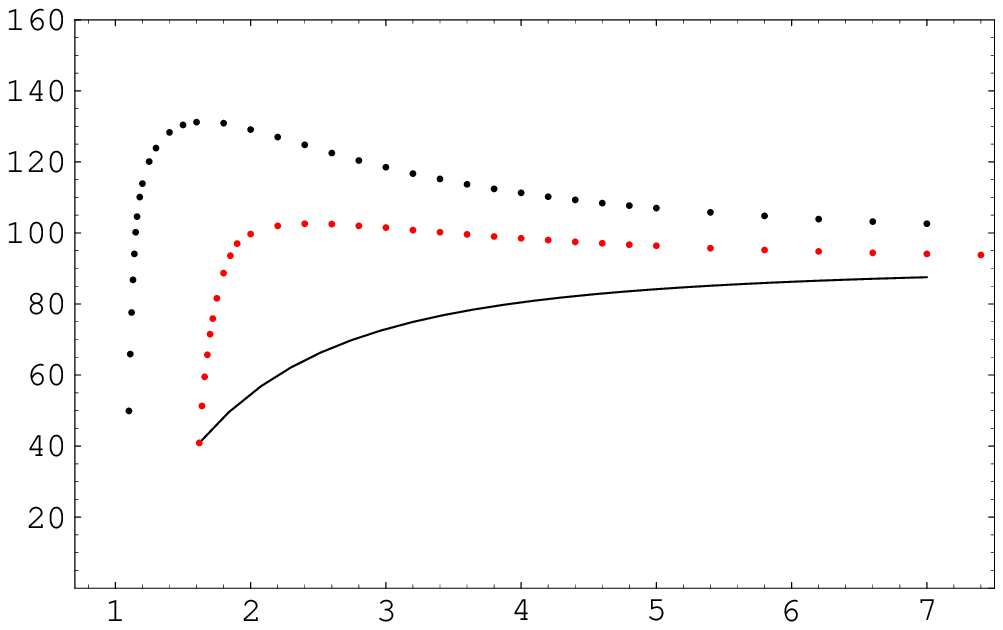}\qquad
\includegraphics[width=80mm]{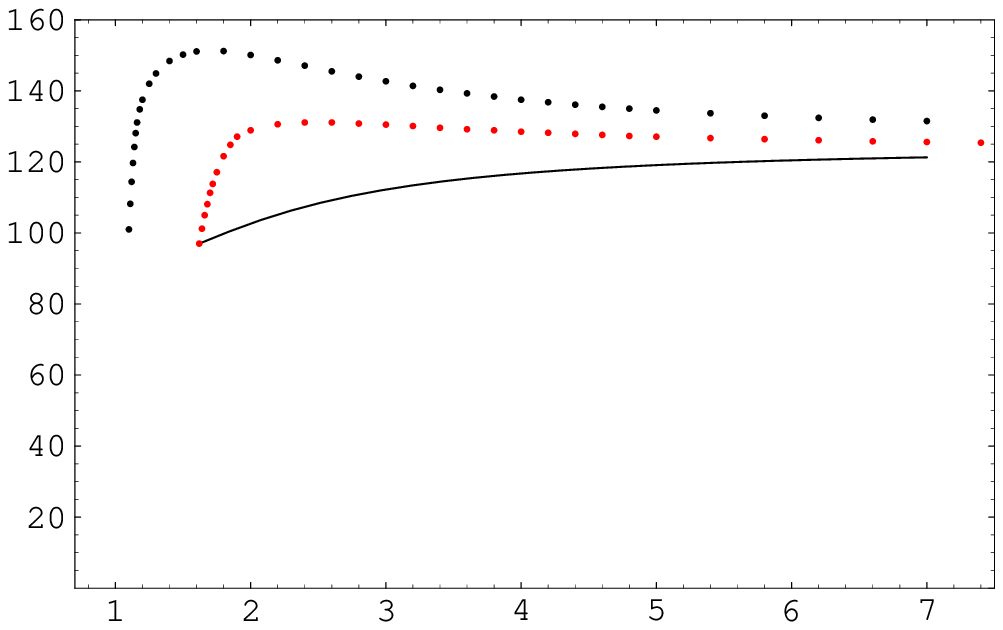}\\
\hspace*{-0.5cm}{$\tan\beta$}\hspace{8cm}{$\tan\beta$}
\caption{(Left) Tree--level upper bound on the lightest Higgs boson mass versus $\tan\beta$.
(Right) The dependence of the two--loop upper bound on the lightest Higgs boson mass on $\tan\beta$ 
for $m_t(M_t)=165\,\mbox{GeV}$, $m_Q^2=m_U^2=M_S^2$, $X_t=\sqrt{6} M_S$ and $M_S=700\,\mbox{GeV}$.
The solid, lower and upper dotted lines represent the theoretical restrictions on the mass of
the lightest CP--even Higgs state in the MSSM, NMSSM and E$_6$SSM respectively.} 
\label{fig2}
\end{figure}

The dependence of the tree level and two-loop upper bounds on the mass of the lightest
Higgs state on $\tan\beta$ is examined in Fig.~\ref{fig2} where these bounds are compared with 
the corresponding limits in the MSSM and NMSSM. One can see that in the interval of $\tan\beta$ 
from $1.2$ to $3.4$ the maximum value of the mass of the lightest Higgs boson in the E$_6$SSM 
is larger than the experimental lower limit on the SM--like Higgs boson even at tree--level. 
At moderate values of $\tan\beta$ ($\tan\beta=1.6-3.5$) the two--loop upper limit on $m_{h_1}$
in the E$_6$SSM is also considerably higher than in the MSSM and NMSSM. It reaches the maximum 
value $\sim 150-155\,\mbox{GeV}$ at $\tan\beta=1.5-2$. In the considered part of the parameter 
space the theoretical restriction on the mass of the lightest CP-even Higgs boson in the NMSSM 
exceeds the corresponding limit in the MSSM because of the extra contribution to $m^2_{h_1}$ 
induced by the additional $F$-term in the Higgs scalar potential of the NMSSM. The size of this
contribution, which is described by the first term in the square brackets of Eq.~(\ref{higgs6}), 
is determined by the Yukawa coupling $\lambda$. The upper limit on the coupling $\lambda$ caused 
by the validity of perturbation theory in the NMSSM is more stringent than in
the E$_6$SSM due to the presence of exotic matter. As a result the upper limit on $m_{h_1}$ 
in the NMSSM is considerably less than in the E$_6$SSM at moderate values of $\tan\beta$.

At large $\tan\beta\gtrsim 10$ the contribution of the $F$-term of the SM-type singlet field 
to $m_{h_1}^2$ vanishes. Therefore with increasing $\tan\beta$ the upper bound on the lightest 
Higgs boson mass in the NMSSM approaches the corresponding limit in the MSSM. In the E$_6$SSM 
the theoretical restriction on the mass of the lightest Higgs scalar also diminishes when 
$\tan\beta$ rises. But even at very large values of $\tan\beta$ the upper limit on $m_{h_1}$ 
in the E$_6$SSM is still $4-5\,\mbox{GeV}$ larger than the ones in the MSSM and NMSSM because
of the $U(1)_{N}$ $D$-term contribution to $m_h$ (last term in the square brackets of 
Eq.~(\ref{higgs6})).

%%%%%%%%%%%%%%%%%%%%%%%%%%%%%%%%%%
\section{Dark Matter and Exotic Higgs decays}
In the E$_6$SSM the lightest SUSY particle tends to be the lightest inert neutralino.
The inert neutralino sector is formed by the neutral components of the inert Higgsinos 
($\tilde{H}^{d0}_1$, $\tilde{H}^{d0}_2$, $\tilde{H}^{u0}_1$, $\tilde{H}^{u0}_2$)
and inert singlinos ($\tilde{S}_1$, $\tilde{S}_2$). In the exact $Z_2^H$ symmetry limit 
the scalar components of the corresponding superfields do not acquire VEVs and 
inert neutralino states do not mix with the ordinary neutralinos. In the field basis 
$(\tilde{H}^{d0}_2,\,\tilde{H}^{u0}_2,\,\tilde{S}_2,\,\tilde{H}^{d0}_1,\,\tilde{H}^{u0}_1,\,\tilde{S}_1)$
the mass matrix of the inert neutralinos takes a form
\begin{equation}
M_{IN}=
\left(
\begin{array}{cc}
A_{22}  & A_{21} \\[2mm]
A_{12}  & A_{11}
\end{array}
\right)\,,\qquad 
A_{\alpha\beta}=-\frac{1}{\sqrt{2}}
\left(
\begin{array}{ccc}
0                                           & \lambda_{\alpha\beta} s           & \tilde{f}_{\beta\alpha} v \sin{\beta}
\\[2mm]
\lambda_{\beta\alpha} s                     & 0                                 & f_{\beta\alpha} v \cos{\beta} \\[2mm]
\tilde{f}_{\alpha\beta} v \sin{\beta}       & f_{\alpha\beta} v \cos{\beta}     & 0
\end{array}
\right)\,,  
\label{icn1}
\end{equation}
so that $A_{12}=A^{T}_{21}$. As before we choose the VEV of the SM singlet field $s$ 
to be large enough ($s > 2400\,\mbox{GeV}$) so that the masses of all inert chargino 
states, which are formed by the charged components of the inert Higgsinos
$(\tilde{H}^{u+}_2,\,\tilde{H}^{u+}_1,\,\tilde{H}^{d-}_2,\,\tilde{H}^{d-}_1)$,
are larger than $100\,\mbox{GeV}$ and $Z'$ boson is relatively heavy. In addition, 
we also require the validity of perturbation theory up to the GUT scale. The 
restrictions specified above set very strong limits on the masses of the lightest 
inert neutralinos. In particular, our numerical analysis indicates that the lightest 
and second lightest inert neutralinos ($\chi^0_{1}$ and $\chi^0_{2}$) are typically 
lighter than $60-65\,\mbox{GeV}$ \cite{10}--\cite{Hesselbach:2007te}. Therefore the 
lightest inert neutralino tends to be the lightest SUSY particle in the spectrum and 
can play the role of dark matter. The neutralinos $\chi^0_{1}$ and $\chi^0_{2}$ are 
predominantly inert singlinos. Their couplings to the $Z$--boson can be rather small 
so that such inert neutralinos would remain undetected at LEP. 

In order to clarify the results of our numerical analysis, it is useful to 
consider a simple scenario when $\lambda_{\alpha\beta}=\lambda_{\alpha}\,\delta_{\alpha\beta}$,
$f_{\alpha\beta}=f_{\alpha}\,\delta_{\alpha\beta}$ and 
$\tilde{f}_{\alpha\beta}=\tilde{f}_{\alpha}\,\delta_{\alpha\beta}$.
In the limit where off--diagonal Yukawa couplings vanish and 
$\lambda_{\alpha} s\gg f_{\alpha} v,\, \tilde{f}_{\alpha} v$ the eigenvalues of the 
inert neutralino mass matrix can be easily calculated (see \cite{Hall:2009aj}).
In particular the masses of two lightest inert neutralino states ($\chi^0_{1}$ and $\chi^0_{2}$)
are given by
\begin{equation}
m_{\chi^0_{\alpha}}\simeq \frac{\tilde{f}_{\alpha} f_{\alpha} v^2 \sin 2\beta}{2\, m_{\chi^{\pm}_{\alpha}}}\,.
\label{icn10}
\end{equation}
where $m_{\chi^{\pm}_{\alpha}}=\lambda_{\alpha} s/\sqrt{2}$ are masses of
the inert charginos. From Eq.~(\ref{icn10}) one can see that the masses of $\chi^0_{1}$ 
and $\chi^0_{2}$ are determined by the values of the Yukawa couplings $\tilde{f}_{\alpha}$ 
and $f_{\alpha}$. They decrease with increasing $\tan\beta$ and chargino masses. 
In this approximation the part of the Lagrangian, that describes interactions of $Z$ with 
$\chi^0_1$ and $\chi^0_2$, can be presented in the following form:
\begin{eqnarray}
\mathcal{L}_{Z\chi\chi}=\sum_{\alpha,\beta}\frac{M_Z}{2 v}
Z_{\mu}\biggl(\overline{\chi^{0}_{\alpha}}\gamma_{\mu}\gamma_{5}\chi^0_{\beta}\biggr)
R_{Z\alpha\beta}\,,\qquad\qquad\qquad\\[0mm]
R_{Z\alpha\beta}=R_{Z\alpha\alpha}\,\delta_{\alpha\beta}\,,\qquad
R_{Z\alpha\alpha}=\frac{v^2}{2 m_{\chi^{\pm}_{\alpha}}^2}
\biggl(f_{\alpha}^2\cos^2\beta-\tilde{f}_{\alpha}^2\sin^2\beta\biggr)\,.
\label{icn14}
\end{eqnarray}
Eqs.~(\ref{icn14}) demonstrates that the couplings of $\chi^0_1$ and $\chi^0_2$ to 
the Z-boson can be very strongly suppressed or even tend to zero. This happens when 
$|f_{\alpha}|\cos\beta\approx |\tilde{f}_{\alpha}|\sin\beta$.

Although $\chi^0_1$ and $\chi^0_2$ might have extremely small couplings to 
$Z$, their couplings to the lightest CP--even Higgs boson $h_1$ cannot be 
negligibly small if $\chi^0_1$ and $\chi^0_2$ have appreciable masses. When the SUSY 
breaking scale $M_S$ and the VEV $s$ of the singlet field are considerably larger
than the EW scale, the lightest CP--even Higgs state is the analogue of the SM Higgs 
field and is responsible for all light fermion masses in the E$_6$SSM. Therefore 
it is not so surprising that in the limit when 
$\lambda_{\alpha} s\gg f_{\alpha} v,\, \tilde{f}_{\alpha} v$ 
the part of the Lagrangian that describes the interactions of $\chi^0_1$ and 
$\chi^0_2$ with $h_1$ takes a form 
\begin{equation}
\mathcal{L}_{H\chi\chi}=\sum_{\alpha,\beta} (-1)^{\theta_{\alpha}+\theta_{\beta}} 
X^{h_1}_{\alpha\beta} \biggl(\overline{\psi^{0}_{\alpha}}(-i\gamma_{5})^{\theta_{\alpha}+
\theta_{\beta}}\psi^0_{\beta}\biggr) h_1\,,\qquad 
X^{h_1}_{\gamma\sigma}\simeq\frac{|m_{\chi^0_{\sigma}}|}{v}\,\delta_{\gamma\sigma}\,,
\label{icn12}
\end{equation}
i.e. the couplings of $h_1$ to $\chi^0_1$ and $\chi^0_2$ are proportional to the mass/VEV. 
In Eq.~(\ref{icn12}) $\psi^0_{\alpha}=(-i\gamma_5)^{\theta_{\alpha}}\chi^0_{\alpha}$ is 
the set of inert neutralino eigenstates with positive eigenvalues, while $\theta_{\alpha}$ 
equals 0 (1) if the eigenvalue corresponding to $\chi^0_{\alpha}$ is positive (negative).

In our analysis we require that the lightest inert neutralino account for all or some of
the observed dark matter relic density. This sets another stringent constraint on the masses
and couplings of $\chi_1^0$. Indeed, because the lightest inert neutralino states are almost
inert singlinos, their couplings to the gauge bosons, Higgs states, quarks (squarks)
and leptons (sleptons) are rather small resulting in a relatively small annihilation
cross section of $\tilde{\chi}^0_1\tilde{\chi}^0_1\to \mbox{SM particles}$ and the possibility 
of an unacceptably large dark matter density. Thus the bulk of the E$_6$SSM parameter space, 
that leads to small masses of $\tilde{\chi}^0_1$, is almost ruled out\footnote{When 
$f_{\alpha\beta},\, \tilde{f}_{\alpha\beta}\to 0$ the masses of $\tilde{\chi}^0_1$ and 
$\tilde{\chi}^0_2$ tend to zero and inert singlino states essentially decouple from the 
rest of the spectrum. In this limit the lightest non-decoupled neutralino may be rather 
stable and can play the role of dark matter \cite{Hall:2011zq}. The presence of very 
light neutral fermions in the particle spectrum might have interesting implications 
for the neutrino physics (see, for example \cite{Frere:1996gb}).}.

A reasonable density of dark matter can be 
obtained for $|m_{\chi^0_{1}}|\sim M_Z/2$ when the lightest inert neutralino states 
annihilate mainly through an $s$--channel $Z$-boson, via its inert Higgsino doublet components 
which couple to the $Z$--boson. If $\tilde{\chi}^0_1$ annihilation proceeds through the 
$Z$--boson resonance, i.e. $2|m_{\chi^0_{1}}|\approx M_Z$, then an appropriate value of 
dark matter density can be achieved even for a relatively small coupling of $\tilde{\chi}^0_1$ 
to $Z$. Since the masses of $\chi^0_1$ and $\chi^0_2$ are much larger than the $b$--quark mass 
and the decays of $h_1$ into these neutralinos are kinematically allowed, the SM--like Higgs
boson decays predominantly into the lightest inert neutralino states and has very small branching
ratios ($2\%-4\%$) for decays into SM particles \cite{10}--\cite{Hall:2010ny}. 

The lightest inert neutralino states can get appreciable masses $\sim M_Z/2$
only if at least one light inert chargino state and two inert neutralinos states, 
which are predominantly components of the $SU(2)_W$ doublet, have masses 
below $200\,\mbox{GeV}$. The inert chargino and neutralinos states that are
mainly inert Higgsinos couple rather strongly to $W$ and $Z$--bosons and therefore can be 
efficiently produced at the LHC and then decay into the LSP and pairs of leptons and quarks 
giving rise to remarkable signatures which can be observed in the near future.

If the masses of $\chi^0_1$ and $\chi^0_2$ are very close then the decays of
$h_1$ into $\chi_{\alpha}\chi_{\beta}$ will give rise to a large invisible branching 
ratio of the SM--like Higgs boson. When the mass difference between the second lightest 
and the lightest inert neutralinos is larger than $10\,\mbox{GeV}$ the invisible branching 
ratio remains dominant but some of the decay products of $\chi_2$ might be observed at 
the LHC. In particular, there is a chance that soft $\mu^{+} \mu^{-}$ pairs may be 
detected. Since the branching ratios of $h_1$ into SM particles are extremely suppressed, 
the decays of the SM--like Higgs boson into $l^{+} l^{-} + X$ could be important for 
Higgs searches \cite{10}.

%%%%%%%%%%%%%%%%%%%%%%%%%%%%%%%%%%
\begin{acknowledgments}
R.N. would like to thank FTPI, University of Minnesota for its hospitality and
M.~A.~Shifman, K.~A.~Olive, A.~I.~Vainshtein, A.~Mustafayev, W.~Vinci, P.~A.~Bolokhov, 
P.~Koroteev for fruitful discussions. Authors are grateful to X.~R.~Tata, M.~Muehlleitner, 
L.~Clavelli, D.~Stockinger, D.~J.~Miller, D.~G.~Sutherland, J.~P.~Kumar, D.~Marfatia,
K.~R.~Dienes, B.~D.~Thomas for valuable comments and remarks. The work of R.N. and S.P. 
was supported by the U.S. Department of Energy under Contract DE-FG02-04ER41291, 
and the work of M.S. was supported by the National Science Foundation PHY-0755262. 
The work of S.M. is partially supported by the NExT Institute. 
S.F.K. and S.M. acknowledge partial support from the STFC Rolling Grant ST/G000557/1.
D.J.M. acknowledges partial support from the STFC Rolling Grant ST/G00059X/1.
J.P.H. is thankful to the STFC for providing studentship funding.
\end{acknowledgments}

\end{document}